\documentclass[twocolumn,pra,showpacs,amsmath,amssymb]{revtex4}
\usepackage{graphicx} \usepackage{epstopdf}
 \usepackage{bm}

\begin{document}

\title{Two-photon and EIT-assisted Doppler cooling in a
three-level cascade system}
 \author{Giovanna Morigi$^1$},
 \author{Ennio Arimondo$^{2}$\footnote{Permanent address: Dipartimento di Fisica
E. Fermi, Universit\`a di Pisa, Italy}}
\affiliation{$^1$ Grup d'Optica, Departament de Fisica, Universitat Aut\`onoma
de Barcelona, 08193 Bellaterra, Spain,\\
 $^2$Laser Cooling and Trapping Group, NIST.
Gaithersburg, MD 20899-8424, USA}

\date{\today}
 \begin{abstract}
Laser cooling is theoretically investigated in a cascade
three-level scheme, where the excited state of a laser-driven 
transition is coupled by a second laser to a top, more stable
level, as for alkali-earth atoms. The second laser action modifies the atomic scattering cross
section and produces temperatures lower than
those reached by Doppler cooling on the lower transition.
When multiphoton processes due to the second
laser are relevant, an electromagnetic induced transparency modifies the absorption of the first laser, and the final
temperature is controlled by the second laser parameters. When the
intermediate state is only virtually excited, the dynamics is
dominated by the two-photon process and the final temperature is determined by the
spontaneous decay rate of the top state. \end{abstract}


 \maketitle

Ultracold temperatures in atomic gases are reached by means of
laser cooling \cite{lasercooling}, evaporative cooling
\cite{evapcooling}, sympathetic cooling \cite{sympcooling} or
stochastic cooling\cite{stochcooling}.  In order to
decrease the kinetic energy associated with the atomic center-of-mass
motion, laser cooling is based on the exchange of
momentum between laser light and a closed atomic (or molecular)
system consisting of a few active levels. In
Doppler cooling the action of light scattering on
the atomic motion can be described by means of a
force~\cite{Nienhuis}. The basic ingredients are i) an atomic
cross section with a resonance enhancing the  photon absorption and
the force acting on the atoms; ii) a
dependence of the absorption process on the atomic momentum
leading to a dependence of the force on the momentum. If the resulting force
damps the atomic momentum, the cooling process
compresses the atomic momenta into a narrow distribution, from
which one extracts the laser cooling temperature.\\
\indent Laser cooling techniques have been demonstrated to be very
efficient for alkali atoms, where temperatures down to several hundreds of
nanoKelvins have been reached~\cite{lasercooling}. Much lower
efficiencies have been achieved for other atomic systems, and in
particular for alkaline-earth-metal atoms. One attractive feature
of group-II elements is the simple internal structure with no
hyperfine levels for the most abundant bosonic isotopes, which
make them particularly attractive for improved frequency standards
and optical clocks. However because the ground state is
nondegenerate, sub-Doppler cooling is not possible and the
temperature for Doppler cooling on the resonance line is typically
limited to a few milliKelvins. For these elements other cooling
strategies have been employed, such as cooling on the
intercombination singlet-triplet line, pioneered in
\cite{katori99}, or quench cooling, modifying the lifetime of a
metastable state by coupling it to a fast-decaying transition,
implemented for neutral calcium atoms in~\cite{binnewies01,curtis01}, and first applied to enhance sideband cooling of ions~\cite{Wineland-Roos}.\\
\indent An alternative cooling strategy 
employs a two-color laser excitation  of a three-level cascade
configuration and is based on mixing by laser radiation the fast decaying singlet
first excited state with another
state having a longer lifetime. This
mixing cooling uses the two-photon atomic coherence created between the ground and top states of the cascade.  Because the Doppler-cooling temperature depends
inversely on the excited state lifetime, a lower temperature may be
produced by the excited state mixing. The scheme
 has flexible handles in the frequency and intensity of
the mixing laser. This scheme was initially explored on
metastable helium with limited efficiency~\cite{kumakura92,rooijakkers97}.
Theoretical analyses on alkali-earth and ytterbium
atoms~\cite{magno03} were followed by experimental evidence of
improved laser cooling for magnesium atoms
\cite{malossi05,moldenhauher06}. These experiments showed that
the temperatures reached by the excited state mixing are
lower than the Doppler temperature determined by the lifetime of
the singlet first excited state, and identified signatures 
hinting to dynamics significantly determined
by the two-photon atomic coherence.\\
\indent In this Rapid Communication we present a theoretical analysis of
laser cooling on a cascade-level scheme with two-color laser
excitation. We determine the laser requirements
for reaching low temperatures, and thereby single out the role of
atomic coherence on the cooling dynamics, identifying the regimes
when it is a key ingredient for reaching efficient cooling. The
atomic level configuration we consider is found in group-II
elements (see the first and second columns in Table~\ref{table1}), and it is
composed by a stable state (ground state), coupled by a laser to
an excited level (intermediate state), which itself is coupled by
a second laser to a higher energy state (top state). The
intermediate state decays more rapidly than the top state, and
the second laser modifies the absorption on the lower transition
by the coupling to the upper transition. This configuration supports
the creation of a stationary atomic coherence between the ground
and top state when the two--photon transition is driven close to
resonance~\cite{Stroud}. \\
\indent Depending on the laser detunings we
identify two different regimes, which characterize the cooling
dynamics. In the first one the formation of atomic coherence
between the ground and top states critically affects the
properties of the force by modifying the lower transition absorption, and temperatures lower than the Doppler
limit on the lower transition are achieved. We denote this regime
as cooling assisted by electromagnetic-induced transparency (EIT),
similar to that realized in a lambda
configuration~\cite{morigi00}. The second regime is characterized
by two-photon processes coupling the ground and  top states,
while the intermediate state is only virtually involved. The
dynamics corresponds thus to that of an effective two-level system
with the linewidth of the upper state. This process produces a
temperature lower than that of the EIT-assisted cooling and
essentially limited by the lifetime
of the upper state. \\
  \begin{table}
 \caption{\label{table1} Parameters for cascade
 atomic transitions.}
 \begin{ruledtabular}
 \begin{tabular}{cccc}
 Parameter & Mg & Ca & Cs\\
 \hline
 $1^{st}$ Trans.& $3^1S_0\rightarrow 3^1P_1$ &  $4^1S_0\rightarrow$ $4^1P_1$ & $6^2S_{1/2}\rightarrow 6^2P_{3/2}$ \\
 $\lambda_{\rm 1}$ (nm) & 285.29  & 422.79  & 852.12 \\
 $\Gamma_{\rm 1}/2\pi$ (MHz) &  78.8 & 34.7 & 5.2 \\
  T$_{\rm D,1}$ (mK) & 1.9  & 0.833  & 0.125 \\
 $2^{nd}$ Trans.& $3^1P_1\rightarrow 3^1D_2$ & $4^1P_1\rightarrow$ $5^1S_0$ & $6^2P_{3/2}\rightarrow 10^2D_{5/2}$ \\
$\lambda_{\rm 2}$ (nm)  &  880.92 & 1034.66 & 563.68\\
 $\Gamma_{\rm 2}/2\pi$ (MHz) &  2.0  & 5.3 \cite{curtis01} & 0.49\cite{neil84}\\
 T$_{\rm D,2}$ ($\mu$K) & 48  & 127  & 12
 \end{tabular}
 \end{ruledtabular}
 \end{table}
\indent The theoretical analysis is based on the time evolution of
the atomic momentum and kinetic energy following the
model discussed in refs.~\cite{wineland79,stenholm86,lett89}. We
restrict our study to one-dimension, and consider an atom of mass
$M$ and momentum $p$ (along the $x$ axis), with internal levels
$|0\rangle,|1\rangle,|2\rangle$ with increasing energies
$0,\hbar \omega_{01},\hbar \omega_{02}$, where $|1\rangle$ ($|2\rangle$)
decays radiatively into $|0\rangle$ ($|1\rangle$) at rate
$\Gamma_1$ ($\Gamma_2$). The Hamiltonian for the atom interacting
with two laser fields at frequencies $\omega_1,\omega_2$ and
wavevectors $k_1,k_2$, respectively, is $H=H_{\rm at}+V_L+W$, where
\begin{eqnarray*}
H_{\rm at}&=&\frac{p^2}{2M}-\hbar\delta_1|1\rangle\langle 1|-
\hbar(\delta_1+\delta_2)|2\rangle\langle2|,\\
V_{L}&=&\hbar\Omega_{1}|1\rangle\langle 0|\cos(k_1x)+\hbar
\Omega_{2}|2\rangle\langle 1|\cos(k_2x)+{\rm H.c.},
\end{eqnarray*} with detunings $\delta_j=\omega_j-\omega_{j0}$
($j=1,2$), Rabi frequencies $\Omega_{1}$ and $\Omega_{2}$.  $W$
describes the coupling to the modes of the electromagnetic field
in the vacuum. We follow the time evolution of an atom, which at
$t=0$ is in the state $|0,p\rangle$ with energy $E_0(p)=p^2/(2M)$,
under the assumption of weak Rabi frequency $\Omega_{1}$, so that
we can treat the coupling of state $|0\rangle$ to state
$|1\rangle$ in perturbation theory~\cite{lounis92}. The scattering
processes causing a change of the atomic momentum are:
\\\indent (i) absorption of one photon with momentum $ \hbar k_1$
along the $x$ axis on the transition $|0\rangle\to|1\rangle$
followed by emission of a photon with momentum $\hbar k_{1s}$. The
final atomic state is $|0,p^{\prime}\rangle=|0,p+\Delta
p_1\rangle$ with $\Delta p_1=\hbar k_1(1-{\hat e}_1\cdot {\hat
x})$ where ${\hat e}_1$ denotes the direction of photon emission.
The corresponding scattering rate is ${\cal W}_{1}(p,{\hat e}_1)=
{\cal P}_1({\hat e}_1){\cal R}_{1}(p)$, where ${\cal P}_1({\hat
e}_1)$ is the spatial pattern of spontaneous emission for
transition $|1\rangle\to |0\rangle$, normalized to unity. The
dependence ${\cal R}_{1}(p)$ on atomic momentum is evaluated using
the approach in~\cite{lounis92}, and takes the value
\begin{equation} {\cal
R}_{1}(p)=\frac{\Gamma_1\Omega_{1}^2}{8}\left|\frac{\delta_1^{\prime}+\delta_2^{\prime}+{\rm
i}\Gamma_2/2}{(\delta_1^{\prime}+\delta_2^{\prime}+{\rm
i}\Gamma_2/2)(\delta_1^{\prime}+{\rm
i}\Gamma_1/2)-\Omega_{2}^2/4}\right|^2 \label{R_1}
 \end{equation} with $\delta_1^{\prime}=- k_1p/M+\delta_1$
 and $\delta_2^{\prime}=-k_2p/M+\delta_2$. The scattering rate for absorption of one photon with momentum $-\hbar k_1$ is analogously evaluated.
 Around the two-photon resonance, $\delta_1+\delta_2=0$, the scattering cross section
 exhibits an asymmetric Fano-like structure as a function of $p$, typical of interference
 processes~\cite{lounis92}. This asymmetry is controlled by $\delta_1$ and
 $\Omega_{2}$, and affects critically the gradient of the cross section at $p=0$, and
thus the force exerted on the atom at slow velocities, as found
in~\cite{morigi00}. \\ \indent (ii) The atom undergoes two
absorption processes at frequencies $\omega_1$ and $\omega_2$
followed by emission of a photon of wave vector $ k_{2s}$ and a
second of wave vector $k_{1s}$ such that the atom is pumped back
to state $|0\rangle$. After this process the atom is found in the
state $|0,p^{\prime}\rangle=|0,p+\Delta p_2 \rangle$ with $\Delta
p_2=\hbar k_1(\pm 1-{\hat e}_1\cdot {\hat x})+\hbar k_2(\pm 1-
{\hat e}_2\cdot\hat{x})$ where ${\hat e}_2$ denotes the direction
of emission of the $k_{2s}$ photon. The scattering
rate for each process is ${\cal W}_{2}(p,{ \hat e}_1,{ \hat e}_2)= {\cal P}_1({\hat
e}_1){\cal P}_2({\hat e}_2){\cal R}_{2}(p)$. Here ${\cal
P}_2({\hat e}_2)$ is the spatial pattern for spontaneous emission
on $|2\rangle\to|1\rangle$ normalized to unity and ${\cal
R}_{2}(p)$ is the convolution of transition
amplitudes, where energy conservation is imposed on the scattering process, and it vanishes when $\Gamma_2$ or $\Omega_2$ are set to zero.  \\
\indent A differential change of the mean kinetic energy $\langle
E\rangle$ is given by the sum of the energy changes due to each
scattering process weighted by their corresponding
rates~\cite{wineland79}. The equation for the evolution of
$\langle E\rangle$  is found taking into account all
absorption-emission paths of copropagating and counterpropagating
photons, and has the form~\cite{note}
 \begin{equation}
\label{energy1}
 \begin{split} &\frac{\rm d}{{\rm d}t}\langle
E\rangle =\int{\rm
d}pf(p)\frac{\hbar p}{M}\Bigl[ k_1\left({\cal R}_{1}(p)-{\cal R}_{1}(-p)\right)\\
&+(k_1+k_2)\left({\cal R}_{2}(p)-{\cal
R}_{2}(-p)\right)\Bigr]+H(\sigma_1,\sigma_2)\end{split}
\end{equation} here $f(p)$ is the momentum distribution and term
 \begin{equation}
 \begin{split}
H(\sigma_1,\sigma_2)&=\sigma_{1}\left(1+\chi_1\right) \frac{\hbar^2 k_1^2}{M}\\
 &+2\sigma_{2}\left(\left(1+\chi_1\right)\frac{\hbar^2k_1^2}{M}
+\left(1+\chi_2\right)\frac{\hbar^2k_2^2}{M}\right),
    \end{split}
    \label{heatingrate}
\end{equation} describes the heating process, with $\chi_i=\int
{\rm d}{\hat e}_i {\cal P}_i({\hat e}_i)({\hat e}_i\cdot{\hat
x})^2$ and $\sigma_i=\int {\rm d}pf(p){\cal R}_{i,}(p)$ ($i=1,2$).
The final temperature is found by considering the latest stages of
cooling, when the Doppler shift is assumed to be smaller than the
natural width of the optical
transitions~\cite{wineland79,stenholm86}. By expanding the rates
${\cal R}_{i}(\pm p)={\cal R}_i(0)\pm{\cal R}_i^{\prime}(0)p$,
${\cal R}_i^{\prime}(0)=d{\cal R}_{i}(p)/dp|_{p=0}$, and $
\sigma_i\sim{\cal R}_i(0)$, we obtain \begin{eqnarray} \frac{\rm
d}{{\rm d}t}\langle E\rangle =-2\alpha\langle E\rangle +H^{\rm o}
\label{energy} \end{eqnarray} where $\alpha$ is the cooling rate
\begin{equation} \alpha=-2\hbar k_1 {\cal R}_1^\prime (0)-2\hbar
\left( k_1+k_2 \right){\cal R}_{2}^\prime(0)
 \end{equation}
 and $H^{\rm o}=H({\cal
R}_1(0),{\cal R}_2(0))$ is the heating rate. The temperature $T$
is operationally defined by the relation $k_{\rm B} T/2= \langle
E\rangle_{\rm st}$, where $k_{\rm B}$ the Boltzmann constant and
$\langle E\rangle_{\rm st}$ is the mean energy, steady state
solution of Eq.~(\ref{energy}). One finds
  \begin{equation}
k_{\rm B} T=H^{\rm o}/\alpha, \label{temperature} \end{equation}
which depends through $H^{\rm o}$ on the parameters $\chi_i$
determined by the spatial pattern of the spontaneous emission and
equal to 2/5 for a dipole emission. For Doppler cooling on the
lower transition the value $\chi_1=1$ reproduces
the temperature reached in three-dimensional cooling~\cite{lett89}.\\
\begin{figure}[t]
\includegraphics[width=6.3cm]{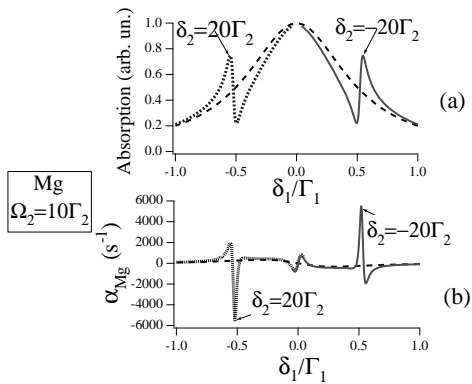}
 \caption{(color online) (a) Absorption coefficient (arbitrary units), and  (b)
cooling rate $\alpha_{\rm Mg}$ versus detuning $\delta_1$ for the
regime of EIT-assisted cooling on the Mg transition in Table
\ref{table1}. The parameters are $\Omega_1=0.01\Gamma_1$ and
$\Omega_2=0$ (dashed line), $\Omega_2=10\Gamma_2$ and
$\delta_2=-20\Gamma_2$ (solid red line), $\Omega_2=10\Gamma_2$ and
$ \delta_2=20\Gamma_2$ (dotted blue line). The parameters $\Omega_2$ and $\delta_2$ correspond to the experimental conditions of \cite{malossi05}. The minima of the absorption
coefficient at $\delta_1 \approx \pm 0.5\Gamma_1$ are at the two-photon
resonance. They are due to atomic coherence between the ground and
top states of the cascade and give rise to a substantial
modification of the cooling rate.}
  \label{MgAbsorpt}
 \end{figure}
 \indent The evaluation of the final temperature $T$ requires the
knowledge of term ${\cal R}_2(p)$, which allows for an explicit
analytical form only in certain limiting cases. Therefore, in our
numerical analysis  we have
linked ${\cal R}_1,{\cal R}_2$ to the steady state solution of the
Optical Bloch Equations for the atomic density matrix
$\rho$~\cite{rooijakkers97}
\begin{equation}
 {\cal
R}_1(p)=\Gamma_1\rho_{11}^{\rm st}(p)-\Gamma_2\rho_{22}^{\rm
st}(p),\,\, {\cal R}_2(p)=\Gamma_2\rho_{22}^{\rm st}(p).
\end{equation}
 where the steady state populations $\rho^{\rm
st}_{\rm ii}$ of the three atomic levels depend on the atomic
momentum $p$ due to the Doppler effect. The results obtained with
this method agree with the prediction obtained using
Eq.~(\ref{R_1}) when $\Gamma_2$ (and thus ${\cal R}_2$) is
zero~\cite{Dalibard}.\\ 
\begin{figure}[ht] \includegraphics[width=6.5cm]{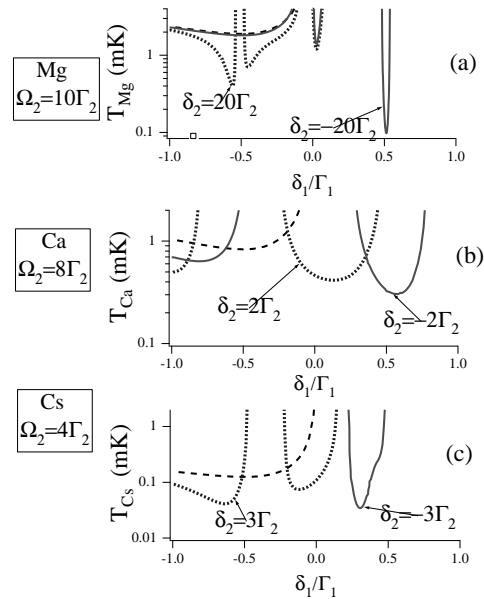}
\caption{(color online) Temperatures  $T$ in mK versus detuning
$\delta_1$ for EIT-assisted cooling of (a) Mg, (b) Ca, and (c) Cs,
using the parameters in Table \ref{table1}, for $\chi_1=\chi_2=1$
as for a three-dimensional cooling geometry. The hyperfine
structure of Cs is not considered. The other parameters are
$\Omega_1=0.01\Gamma_1$, where $\Gamma_1$ depends on the atomic
species, $\Omega_2$ and $\delta_2$ are specified in the figures.
The continuous red (dotted blue) lines indicate the temperatures
obtained for negative (positive) values of $\delta_2$, the dashed
line shows the Doppler cooling temperature obtained when
$\Omega_2=0$. }
 \label{MgCaCsTemp}
 \end{figure}
We have evaluated the cooling temperatures as a function of laser
and atomic parameters for the different species listed in Table
\ref{table1}. $T_{\rm D1,2}$ denotes  the Doppler cooling limit
temperature $\hbar \Gamma_{1,2}/(2 k_{\rm B})$ for a two-level
system with the upper lifetime equal to that of the $|1\rangle$ or
$|2\rangle$ state, respectively. In order to simplify the
theoretical treatment we treat all optical transitions as closed.\\
\indent Figure \ref{MgAbsorpt}(a) displays the absorption
coefficient of Mg atoms (where $\Gamma_1/\Gamma_2 \approx 39$) for
Doppler cooling on the lower transition (dashed line) and for
EIT-assisted cooling with a strong near resonant coupling laser on
the upper transition (continuous and dotted lines). The coupling
laser modifies strongly the lower transition absorption profile
around $\delta_1+\delta_2\approx 0$. For the  chosen parameters
the absorption profile contains a narrow and deep structure at
positive or negative detuning $\delta_1$, depending on the sign of
$\delta_2$. In presence of the coupling laser, the cooling rate
$\alpha_{\rm Mg}$, proportional to the derivative of the
absorption coefficient with respect to the frequency, presents
deep and narrow structures, as in Fig. \ref{MgAbsorpt}(b). The
larger values of $\alpha$ in Fig.~\ref{MgAbsorpt}(b) correspond to
lower cooling temperatures, as shown in Fig. \ref{MgCaCsTemp}(a),
see also Eq. (\ref{temperature}). For
$\delta_2>0$ the EIT process increases the cooling rate by twenty
times, and reduces the Doppler limit temperature by a factor of
five. Figures~\ref{MgCaCsTemp}(b) and~(c) display similar
behaviors for the Ca and Cs transitions. The Ca case confirms a
peculiar feature of the Mg results: the lowest temperatures are
reached about the two-photon resonance for positive detuning
$\delta_1$, corresponding to the largest increase in the damping
rate as shown in Fig. \ref{MgAbsorpt}(b). A similar dependence was
already observed in the EIT cooling of~\cite{morigi00}. EIT-assisted
cooling at positive $\delta_1$ is difficult to realize because its
efficiency is sensitive to changes of $\Omega_2$, which may occur
along the atomic sample: in regions where $\Omega_2\sim0$ the
atoms are heated by the laser coupling to the lower transition.
For the explored parameters of Mg and Ca the temperatures reached in this EIT-assisted cooling are not very close to the Doppler cooling limit on the upper transition. Interestingly, in Cs low temperatures are achieved also using a laser at short wavelength on the upper transition, corresponding to a large value of $k_2$ and contributing to an increase the last term in the heating rate of Eq. (\ref{heatingrate}).  \\
\begin{figure}[ht] \includegraphics[width=9cm]{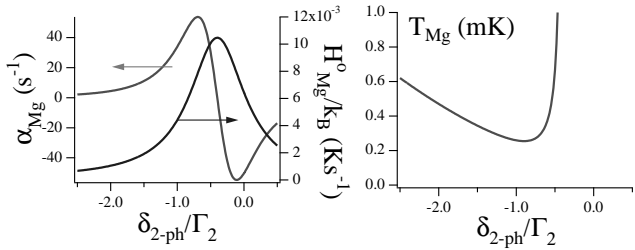}
\caption{Two-photon cooling of Mg. (Left) Cooling rate
$\alpha_{\rm Mg}$ and heating rate $H^{\rm o}_{\rm Mg}$, and
(right) temperature $T_{\rm Mg}$ (right), versus the 2-photon detuning $\delta_{\rm
2-ph}=\delta_1+\delta_2$. The parameters are $\Omega_1=0.01
\Gamma_1$, $\delta_1=-40\Gamma_1 $ and $\Omega_2=50\Gamma_2$. }
 \label{MgRaman}
 \end{figure}
\indent Tuning the intermediate state $|1\rangle$ far-off
resonance, while keeping the two lasers close to two-photon
resonance, one switches from the EIT-assisted cooling regime,
where atomic coherence plays an important role on the cooling
dynamics, to the two-photon cooling regime, where the dynamics is
essentially described by an effective two-level system composed by
the states $|0\rangle$ and $|2\rangle$. Here, cooling is expected
to produce a temperature close to the Doppler limit $T_{\rm D2}$
on the upper level. This is confirmed by the numerical analysis
displayed in Fig. \ref{MgRaman} for Mg, where the $\omega_1$ laser
is far detuned by 40 linewidths from the $|0\rangle \to |1\rangle$
transition and the frequency sum $\delta_{\rm
2-ph}=\delta_1+\delta_2$ is scanned over a small frequency
interval determined by the $\Gamma_2$ decay rate. Temperatures
$T\sim T_{\rm D2}$, and much smaller than $T_{\rm D1}$, are
obtained for different values of $\delta_1$ and $\Omega_2$. The
application of this scheme requires a control of the laser
sum frequency $\delta_1+\delta_2$ with a precision
determined by $\Gamma_2$, which is challenging in the case of an
ultraviolet laser as required for Mg. An analysis of the velocity
capture range $\Delta v_{\rm c}$ shows that a lower temperature is
reached at the expense of the lower efficiency in the collection
of the cooled atoms. This can be seen by comparing the capture
range of the EIT process $\Delta v_{\rm c}\sim 0.2\Gamma_1/k_1$,
as derived from the frequency range of Fig. \ref{MgAbsorpt} where
the the damping rate is significantly different from that of the
Doppler cooling process, with the the velocity capture range of
two-photon cooling $\Delta v_{\rm c}\sim \Gamma_2/k_2$, as derived
from Fig. \ref{MgRaman}. Moreover, the numerical results point
also out that in EIT-assisted cooling the cooling rate $\alpha$ is
much larger than in the two-photon
cooling.\\
\indent In conclusion, we have analysed laser cooling based on
two-color excitation of an atomic cascade, when the upper state
has a longer lifetime. This scheme is complementary to that of
quenching cooling where the stability of top and intermediate
states is exchanged. We have discussed two regimes, EIT-assisted
and two-photon cooling. Lower temperatures may be reached using
two-photon excitations, with the intermediate state excited only
virtually, while in the EIT-assisted scheme a larger capture range
and a more flexible combination with the
single photon Doppler cooling enhance the cooling performance.\\
\indent One of the author (E.A.) wishes to thank C. Oates, J.V.
Porto and J. Thomsen for useful discussions. G.M. acknowledges
support of the Harvard Smithsonian Center for Theoretical Atomic,
Molecular, and Optical Physics during completion of this work.


\begin{thebibliography}{99}
\bibitem{note} Within Eq.~(\ref{energy1}) we have neglected small
terms depending on $k_1-k_2$ in the damping coefficient, and a
term associated with the correlation in the spontaneous emission
of photons on the two atomic transitions. This latter term is
expected to have an amplitude similar to those containing the
$\chi_{\rm i}$ parameters, and should
increase the final temperature by less than fifty percent.
\end{thebibliography}
 \end{document}